# The risk of bias in denoising methods

Kendrick Kay[1]*

[1]Center for Magnetic Resonance Research (CMRR), Department of Radiology, University of Minnesota
*Corresponding author (kay@umn.edu)

# Abstract

Experimental datasets are growing rapidly in size, scope, and detail, but the value of these datasets is limited by unwanted measurement noise. It is therefore tempting to apply analysis techniques that attempt to reduce noise and enhance signals of interest. In this paper, we draw attention to the possibility that denoising methods may introduce bias and lead to incorrect scientific inferences. To present our case, we first review the basic statistical concepts of bias and variance. Denoising techniques typically reduce variance observed across repeated measurements, but this can come at the expense of introducing bias to the average expected outcome. We then conduct three simple simulations that provide concrete examples of how bias may manifest in everyday situations. These simulations reveal several findings that may be surprising and counterintuitive: (i) different methods can be equally effective at reducing variance but some incur bias while others do not, (ii) identifying methods that better recover ground truth does not guarantee the absence of bias, (iii) bias can arise even if one has specific knowledge of properties of the signal of interest. We suggest that researchers should consider and possibly quantify bias before deploying denoising methods on important research data.

# Introduction

Modern science has witnessed major advances in the application of computational analyses to large datasets (LeCun et al., 2015; Marx, 2013). This has led to a 'big data' revolution in which datasets of increasing size, scope, and detail are being amassed (Abbott, 2021; Blei and Smyth, 2017; Deng et al., 2009). In the field of neuroscience, advances in electrophysiological, optical, and magnetic resonance techniques are enabling measurement of the structure and function of animal and human brains at higher resolution, with greater coverage, and over longer temporal durations. However, a major challenge in these measurements is the presence of noise, which we define as unwanted variability across repeated measurements from the same individual. Such noise can originate from a variety of sources and can be both structured (e.g., imaging artifacts, head motion, physiological noise, variations in cognitive performance) and unstructured (e.g., thermal noise, optical shot noise). Depending on the goals of a given experiment, many of these types of noise are undesirable to the researcher.

Developing methods for removing noise from data has been a long-standing objective in neuroscience. High levels of noise in experimental data hinder scientific inferences; thus, there is a temptation to apply denoising methods to such data. Indeed, there are many interesting recently proposed approaches for denoising, including low-rank methods (Mason et al., 2021; Veraart et al., 2016; Vizioli et al., 2021), methods based on data-driven noise derivation (Behzadi et al., 2007; Pruim et al., 2015; Salimi-Khorshidi et al., 2014), methods that exploit the power of deep neural networks (Knoll et al., 2020; Lecoq et al., 2021; Qiao et al., 2021; Yang et al., 2020), and self-supervised methods (Fadnavis et al., 2020). In surveying the literature, we find extensive discussion and consideration of denoising methods and how they fare in specific scientific paradigms. However, we think that, aside from a few notable exceptions (Huang et al., 2021; Lerma-Usabiaga et al., 2020), there has been insufficient emphasis on the issue of statistical bias.

Bias, in the statistical sense, is defined as the discrepancy between the average expected outcome of a given experiment (and its associated analysis) and the ground-truth parameter being estimated (a more formal treatment is provided later). In expositions of denoising methods, bias is often not even mentioned, let alone quantified and assessed. Coming to clarity on this methodological issue is especially important in the context of modern datasets. This is because increasing sizes of datasets, increasing levels of noise (due to increased spatial resolution, temporal resolution, and acquisition speeds), and increasing complexity of data analysis pipelines all tend to obscure or make more difficult the assessment of bias. A critical message of this paper is that bias is risky: while a method might improve the correspondence between a noisy dataset and a ground-truth measure, this might come at the cost of introducing systematic biases into the data and lead to incorrect scientific inferences.

We write this article with two goals in mind. First, we wish to draw attention to—or perhaps rekindle interest in—the basic statistical concepts of bias and variance. Our presentation is general in order to isolate the essential principles at stake. Second, we wish to communicate several simulations that illustrate how these concepts and principles can be applied in concrete scientific paradigms. We design these examples based on our experience in neuroimaging, and we make freely available the underlying data and code to promote transparency (files available at https://osf.io/weg87/). The examples are not intended to establish general methodological findings (for that, more extensive analyses are necessary), but rather to provide important insights into the nature of denoising. We acknowledge that the ideas and principles we convey may already be apparent to expert practitioners. Thus, perhaps the primary audience of this paper are researchers who are interested in—but have not fully developed their stance towards—strategies for denoising data. Ultimately, we hope this article spurs method developers to consider and potentially quantify bias in candidate denoising methods and users to consider the risk of bias when applying denoising methods to important research data.

## Methods

Simulation framework

All simulations (as depicted in **Figures 2–4**) use a common analytical framework. We first design a ground truth based on either empirical or synthetic data. We then generate simulated data by adding randomly generated noise to the ground truth. This produces a set of measurements, each of which may contain multiple data points (e.g. different voxels, different time points). Next, we apply various denoising methods. Each method is applied independently to each measurement and produces a set of analysis results. Finally, for each method, we compute quantitative metrics that assess the performance of the method. Three metrics are computed and are detailed below.

*Bias* is quantified by computing, for each data point, the absolute deviation between the mean across analysis results and the ground truth, normalized by the standard error across analysis results (this normalization can be viewed as a form of studentization). Note that computing the absolute value is important, since a denoising method might overestimate and underestimate the ground truth in different parts of a dataset and it should be penalized for doing so. We summarize the results by calculating the median deviation across data points. The values are in normalized units, and low values are desirable, as they indicate low deviations from ground truth. Data points for which the standard error across analysis results is 0 are ill-defined and are ignored in the calculation (e.g. **Figure 3B, right column, time = 0 s**). It is important to note that our metric of bias is not, strictly speaking, the same as the idealized theoretical definition of statistical bias (see **Equation 1**). The theoretical definition would require computing expectation over an infinite (or very large) number of simulations; in contrast, our metric is suitable for computation in finite data regimes and takes into account the limited number of simulations through normalization by standard error. One issue with the metric is that non-zero values are obtained even for unbiased measurements (thus, the metric can be viewed as the "apparent bias"). Therefore, to provide a suitable comparison, we perform Monte Carlo simulations (assuming a Gaussian noise distribution) to determine the value that is expected for the case of unbiased measurements; this value is approximately 0.70 and is plotted as 'Baseline' in **Figures 2–4**.

*Variance* is quantified by computing, for each data point, the standard error across analysis results. We summarize the results by calculating the median standard error across data points. The values are in the units of the original data, and low values are desirable, as they indicate high reliability of analysis results.

*Error* is quantified by computing Pearson's correlation between each analysis result and the ground truth. (Note that correlation allows flexibility for scaling and offset; while a non-flexible metric such as mean squared error is technically more correct, correlation is appealing for its interpretable units and is likely sufficient in most cases.) We summarize the results by calculating the mean correlation observed across analysis results. Intuitively, this metric assesses how well a denoising method recovers ground truth. Correlation values range from –1 to 1. High values are desirable, as they indicate high similarity of analysis results to the ground truth.

Simulation 1: Anatomical data

In this simulation, we use as ground truth the pre-processed 0.8-mm $T_1$-weighted anatomical volume acquired from Subject 1 from the Natural Scenes Dataset (NSD) (Allen et al., 2022). The intensity values in this volume range approximately from 0 to 1400 (see **Figure 2A, middle**). Also from NSD, we use the brain mask calculated for the subject and the tissue segmentation provided by FreeSurfer (see **Figure 2A, bottom**). We map the 1-mm MNI $T_1$-weighted atlas provided with FSL (https://fsl.fmrib.ox.ac.uk/fsl/) to the subject-native anatomical space using linear interpolation (see **Figure 2A, top**). We generate a set of 10 measurements by adding noise drawn from a Gaussian distribution with mean zero and standard deviation 300 (noise drawn independently for each voxel). We evaluate four denoising methods: (1) *No denoising* refers to using the measurements as-is. (2) *Gaussian smoothing* refers to spatially smoothing a given measurement using a 3D isotropic Gaussian kernel with a full-width-half-maximum (FWHM) of 3 mm. (3) *MNI atlas prior* refers to averaging a given measurement with the MNI atlas (mapped to subject-native space). Before averaging, a scale and offset is applied to the atlas such that the mean of the data within gray matter (as indicated by the tissue segmentation) and the mean of the data within white matter are matched to the corresponding gray- and white-matter means in the MNI atlas. (4) *Anisotropic smoothing* refers to applying nonlinear anisotopic diffusion-based smoothing (Weickert, 1998) as implemented in Segmentator (Gulban et al., 2018). The diffusion-based smoothing is run for 20 iterations. For all denoising methods, quantitative metrics of performance (as described previously) are computed using voxels within the brain mask.

Simulation 2: Response timecourses

In this simulation, we use as ground truth a synthetic hemodynamic response function (HRF) generated by evaluating a double-gamma function as implemented in SPM's spm_hrf.m (https://www.fil.ion.ucl.ac.uk/spm/). The parameters [6 16 1 1 2 0] are used; these are the defaults, except for the fifth parameter, which is set to create a strong undershoot. The double-gamma function is convolved with a 1-s boxcar, sampled at a rate of 1 s, and then scaled to peak at 1. The resulting HRF represents a hypothetical fMRI response timecourse to a 1-s stimulus (see **Figure 3A, top**). We generate a set of 10 measurements by adding temporally correlated Gaussian noise with mean zero and standard deviation 0.2 (this was accomplished by generating zero-mean Gaussian noise with standard deviation 0.2 and convolving the noise with a 5-s boxcar with integral 1). We evaluate three denoising methods: (1) *No denoising* refers to using the measurements as-is. (2) *Basis restriction* refers to projecting the measurements onto a set of basis functions and then reconstructing the measurements. For basis functions, we take the library of 20 canonical HRFs obtained from the Natural Scenes Dataset (Allen et al., 2022) (getcanonicalhrflibrary.m), predict the response to a 1-s stimulus, perform principal components analysis on the 20 timecourses, and extract the top three principal component timecourses (see **Figure 3A, bottom**). (3) *Parametric fit* refers to fitting each measurement using a double-gamma model (same as used to generate the data). Specifically, we use nonlinear optimization (MATLAB Optimization Toolbox's lsqnonlin.m) to determine the optimal parameters for a double-gamma function (as implemented in SPM's spm_hrf.m) such that when convolved with a 1-s boxcar, the result best approximates the measurement in a least-squares sense. The initial seed for the optimization is set to [6 16 1 1 6 0], which are the defaults in spm_hrf.m.

Simulation 3: Tuning curves

In this simulation, we use as ground truth a synthetic set of tuning curves associated with several hypothetical units (these units can be thought of as individual neurons or voxels). We construct tuning curves that represent the response of 10 units to 50 conditions—these conditions can be viewed as different points along some hypothetical stimulus dimension. We fix the dimensionality of the representation to be exactly 4. This is accomplished by creating 4 Gaussian functions spaced equally along the stimulus dimension, and then generating tuning curves for each unit by weighting and summing these Gaussian basis functions using a set of randomly generated weights (random numbers are drawn from a uniform distribution between 0 and 1 and then cubed). Each unit's tuning curve is scaled to peak at 1, and to aid visibility, units are arranged in sorted order according to the center-of-mass of each tuning curve (see **Figure 4A**). We generate a set of 30 measurements by adding noise drawn from a Gaussian distribution with mean zero and standard deviation 0.6. (For visibility, only 10 of these 30 measurements are shown in **Figure 4B, bottom row**.) We evaluate three denoising methods. (1) *No denoising* refers to using the measurements as-is. (2) *Boxcar smoothing* refers to smoothing each unit's measured tuning curve using a boxcar kernel with width 3 and integral 1 (this is simply a moving average with window size 3). (3) *PCA* refers to reducing the dimensionality of each measurement to a specific target rank, a method also referred to as Truncated SVD (Gavish and Donoho, 2017). Variants of this method can be found in the literature (e.g. Veraart et al., 2016; Vizioli et al., 2021). Specifically, given a measurement $X$ (10 units × 50 conditions), we perform singular value decomposition to obtain $X = USV^T$ where $U$ (10 × 10) has loadings in the columns, $S$ (10 × 50) has singular values in decreasing order on the diagonal and zeros elsewhere, and $V$ (50 × 50) has timecourse components in the columns. We then perform low-rank reconstruction of the measurement using $n$ = 2, 3, 4, 6, or 8 components (referred to as PCA2, PCA3, PCA4, PCA6, and PCA8) by computing the reconstructed measurement $X^* = U^*S^*V^{*T}$ where $U^*$ contains the first $n$ columns of $U$, $S^*$ contains the upper-left $n \times n$ elements of $S$, and $V^*$ contains the first $n$ columns of $V$.

# Results

## A brief review of bias and variance

We start by briefly reviewing some basic statistical concepts (Hastie et al., 2001; Raunig et al., 2015). Suppose we are interested in estimating a certain population parameter by performing measurements of this parameter. There are two distinct aspects of the quality of our measurements: bias and variance. *Bias* refers to the discrepancy, if any, between the average expected outcome of our measurements and the population parameter. All else being equal, we want bias to be zero (or low), since we want our measurements to cluster around the true value of the population parameter. *Variance* refers to the variability of our measurements. All else being equal, we want variance to be low, since this helps us narrow down a range of plausible values for the population parameter.

   A simple example helps illustrate these concepts. **Figure 1** depicts a 2 × 2 crossing of different measurement scenarios. The columns differ in the amount of measurement bias. The left column corresponds to unbiased measurement, in which measurements, on average, equal the ground-truth parameter, whereas the right column corresponds to biased measurement, in which measurements, on average, are higher than the ground-truth parameter. The rows differ in the amount of measurement variance. The top row corresponds to low-variance measurement, in which measurements cluster tightly together, whereas the bottom row corresponds to high-variance measurement, in which measurements are spread far apart.

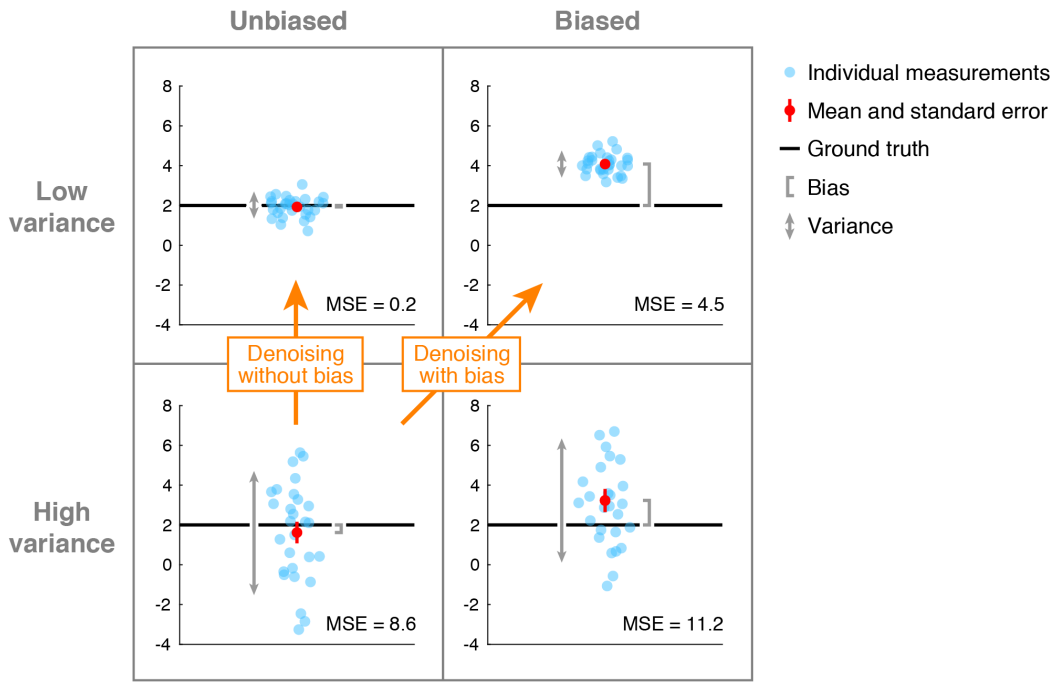

**Figure 1. Bias and variance in measurement.** In each of the four depicted simulations, 2 is the ground-truth value and 30 measurements are simulated by drawing values from a Gaussian distribution. In the left column, the Gaussian distributions have a mean of 2 (unbiased), whereas in the right column, the distributions have a mean of 4 (biased). In the top row, the Gaussian distributions have a variance of 0.3 (low variance), whereas in the bottom row, the distributions have a variance of 8 (high variance). The inset indicates the mean squared error (MSE) between the measurements and the ground truth. Bias can be estimated as the discrepancy between the mean of the measurements and the ground truth. Variance can be estimated as the variability across the measurements. Code available at https://osf.io/6x8kq/.

A common approach for assessing how well a measurement procedure captures the population parameter is to compute *mean squared error* (MSE), which refers to the average squared deviation of the measurements from the population parameter. It is important to note that this error metric reflects *separate contributions of bias and variance*. Specifically, mean squared error is equal to the sum of two separate terms, a squared-bias term and a variance term:

$$\text{MSE} = \text{BIAS}^2 + \text{VARIANCE} \qquad [1]$$

To see why this is the case, we first define bias as the difference between the average expected measurement and the ground-truth value:

$$\text{BIAS} = \mathbb{E}[\hat{y}] - y \qquad [2]$$

where $y$ indicates the ground-truth value, $\hat{y}$ indicates a single measurement, and $\mathbb{E}$ is the expectation operator indicating the average over an infinite number of repeated measurements. We compute the squared bias as follows:

$$\text{BIAS}^2 = y^2 - 2y\mathbb{E}[\hat{y}] + (\mathbb{E}[\hat{y}])^2 \qquad [3]$$

Note that squared bias is always non-negative. Next, we define variance as the average squared deviation of the measurements around their mean:

$$\begin{aligned}\text{VARIANCE} &= \mathbb{E}[(\hat{y} - \mathbb{E}[\hat{y}])^2] \\ &= \mathbb{E}[\hat{y}^2 - 2\hat{y}\mathbb{E}[\hat{y}] + (\mathbb{E}[\hat{y}])^2] \\ &= \mathbb{E}[\hat{y}^2] - 2\mathbb{E}[\hat{y}]\mathbb{E}[\hat{y}] + (\mathbb{E}[\hat{y}])^2 \\ &= \mathbb{E}[\hat{y}^2] - (\mathbb{E}[\hat{y}])^2 \qquad [4]\end{aligned}$$

Finally, we define mean squared error as the average squared deviation of the measurements from the ground-truth value:

$$\begin{aligned}\text{MSE} &= \mathbb{E}[(y - \hat{y})^2] \\ &= \mathbb{E}[y^2 - 2y\hat{y} + \hat{y}^2] \\ &= y^2 - 2y\mathbb{E}[\hat{y}] + \mathbb{E}[\hat{y}^2] \qquad [5]\end{aligned}$$

Adding some terms and grouping, we obtain:

$$\text{MSE} = (y^2 - 2y\mathbb{E}[\hat{y}] + (\mathbb{E}[\hat{y}])^2) + (\mathbb{E}[\hat{y}^2] - (\mathbb{E}[\hat{y}])^2) \qquad [6]$$

By substituting from **Equations 3 and 4**, we see:

$$\text{MSE} = \text{BIAS}^2 + \text{VARIANCE} \qquad [7]$$

Insights and implications for denoising

Having reviewed the concepts of bias and variance, we highlight some important insights. First, we remind ourselves of the classic distinction between reliability and accuracy. Even though a procedure might provide highly reliable measurements (low variance), this does not necessarily imply that that the measurements are accurate. This is because the measurements might have systematic deviation (bias) from the ground-truth parameter (e.g., see upper-right panel of **Figure 1**). Second, we observe that assessing error relative to ground truth does not provide specific information regarding bias. Error, as discussed earlier, reflects the combination of both bias and variance. Hence, a situation in which error is low is compatible with the existence of bias (e.g., in **Figure 1**, the upper-right panel exhibits lower error than the lower-left panel but has substantial bias).

We now transition to the topic of denoising. A common situation that an experimentalist may face is one in which a set of measurements are corrupted by high levels of noise but are at least expected to converge, across repeated experiments, to the true signal. This situation can be characterized as high variance and unbiased (**Figure 1, lower left**). To reduce noise, the experimentalist might try applying a denoising technique to the data. In doing so, there are two general types of outcomes. One outcome is that variance is reduced while the absence of bias is maintained (see arrow labeled 'Denoising without bias' that begins in the lower-left panel and ends in the upper-left panel). This is a great outcome. A different outcome is that variance is reduced but bias is introduced (see arrow labeled 'Denoising with bias' that begins in the lower-left panel and ends in the upper-right panel). This is arguably an undesirable outcome, as repeated experiments converge to an incorrect signal. Reduction of variance but introduction of bias is an instance of the classic bias-variance tradeoff (Hastie et al., 2001).

Examples of bias and variance in denoising

While we have described theoretical considerations to take into account when assessing a denoising method, it may be unclear how much these considerations actually matter in practical situations. To provide more concrete insights, we construct three denoising simulations based on our experience with neuroimaging data. The goal of these simulations is to provide examples of how the performance of different denoising methods can be formally evaluated. In each example, we start with a ground truth, generate noisy measurements based on this ground truth, apply different denoising methods to each measurement, and calculate metrics that quantify the performance of the denoising methods. We generally follow the theory presented earlier, but use versions of the metrics that are more suitable and interpretable for practical data scenarios. Specifically, we quantify *bias* as the median absolute deviation between the mean across analysis results and the ground truth and express this in units of standard error; we quantify *variance* as the median standard error across analysis results; and we quantify *error* as the average correlation between each analysis result and the ground truth (see Methods). Please note that the denoising methods demonstrated in the examples are not intended to be realistic methods that one might want to use in practice (e.g., Gaussian smoothing is obviously a naive approach; averaging with an MNI atlas is obviously a very crude approach). This is because the point of the examples is not so much to determine the best state-of-the-art denoising method, but rather to demonstrate how bias and variance can be formally studied.

In the first simulation, we use as ground truth a high-quality 0.8-mm isotropic anatomical MRI scan of a human brain (**Figure 2A**) and simulate noisy measurements of this ground truth by adding Gaussian noise. As expected, the raw data ('No denoising') follow the ground truth, in the sense of lacking bias, but suffer from high variance (**Figure 2B, first column**). The method of spatial smoothing ('Gaussian smoothing') reduces variance, but incurs major deviations from ground truth (**Figure 2B, second column**). This is not surprising since the smoothing kernel used has a relatively large full-width-half-maximum of 3 mm, which will obviously remove fine-scale features of the convoluted cerebral cortex. The method of averaging a given measurement with a pre-existing atlas ('MNI atlas prior') provides some variance reduction, but also introduces some bias (**Figure 2B, third column**). This makes sense, since the atlas is generally expected to provide good guesses for tissue intensity, but may bias the measurement in parts of the individual's brain that deviate from the atlas. Finally, the method of applying anisotropic smoothing ('Anisotropic smoothing') greatly reduces variance and, appealingly, introduces very little bias, if any (**Figure 2B, fourth column**). Our interpretation is that the assumption embodied by anistropic smoothing—namely, that true structures are locally contiguous and have homogeneous signal intensity—is well matched to the anatomical structure of the brain, at least at the current spatial resolution.

The quantitative summary plots (**Figure 2C**) provide interesting insights. Anisotropic smoothing reduces variance but does not incur appreciable bias (arrow 1). In contrast, other methods such as Gaussian smoothing reduce variance but incur substantial bias (arrow 2). Thus, a bias-variance tradeoff does not necessarily occur in all situations. We also see that error is not a perfect metric to discriminate amongst methods, as both anisotropic smoothing (location 3) and Gaussian smoothing (location 4) yield comparable levels of correlation between analysis results and ground truth. Finally, there is a general relationship between reducing variance and increasing similarity to ground truth (arrow 5). This makes sense since denoising methods should, in theory, reduce unwanted measurement noise and generally push results towards the ground truth.

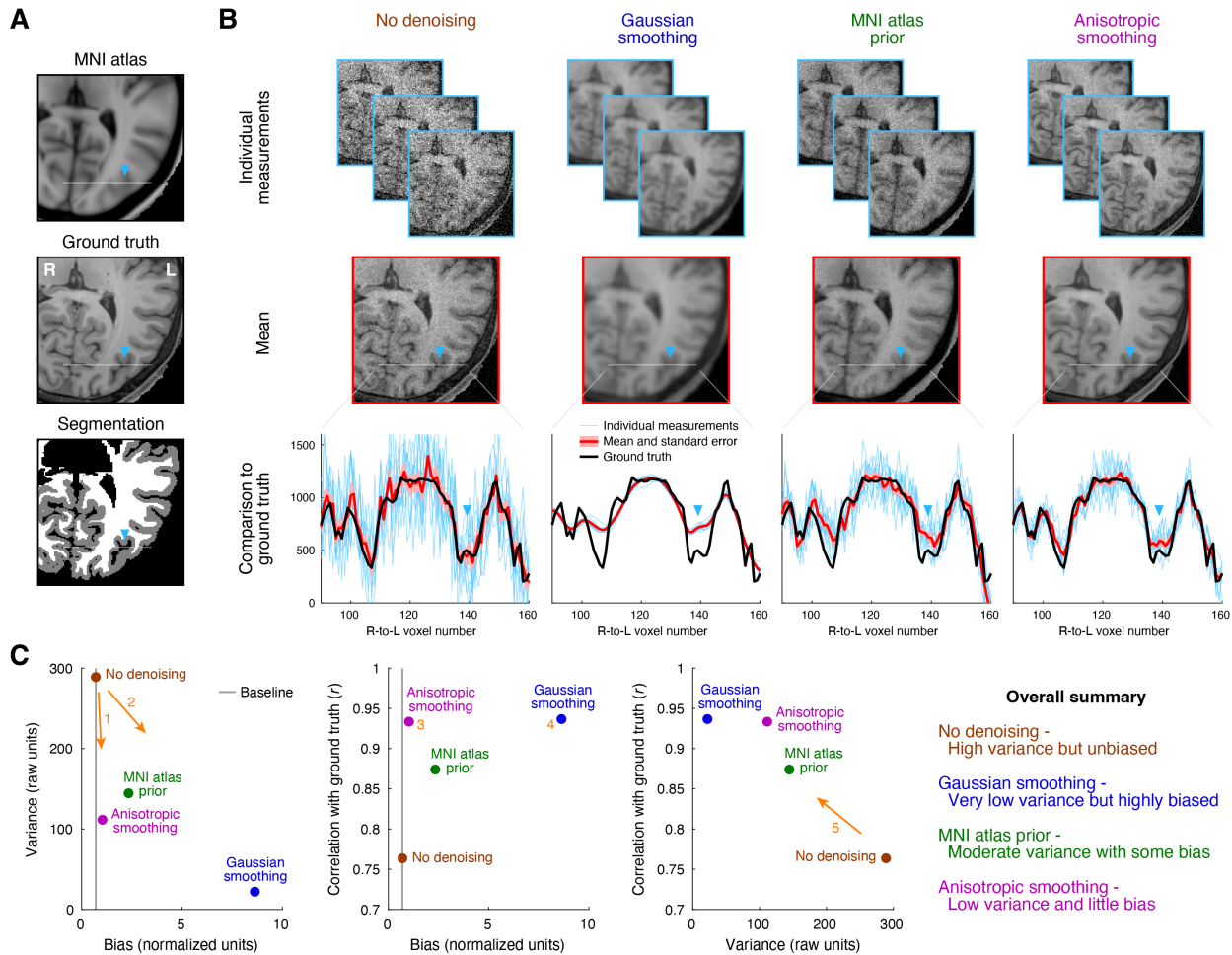

**Figure 2. Denoising anatomical data.** In this simulation (code available at https://osf.io/qxp8y/), we generate noisy measurements by starting with a ground-truth $T_1$-weighted anatomical volume and adding Gaussian noise independently to each voxel. We then attempt to denoise the data using different denoising methods: no denoising, simple Gaussian spatial smoothing, averaging with a group-average atlas prior, and performing anisotropic diffusion. The images depict a zoomed-in view of the posterior section of a single axial slice, and the same color map and range is used for all images. *A*, Reference volumes. We illustrate the ground-truth anatomical volume (middle), the MNI atlas used in one of the denoising methods (top), and the tissue segmentation obtained from FreeSurfer, showing gray and white matter (bottom). *B*, Denoising results. Each column shows results for a different denoising method. We show three example measurements (top row), the mean across measurements (middle row), and detailed plots for a small line of voxels (bottom row). *C*, Quantitative assessment of bias, variance, and error. Bias is quantified as the median absolute difference between the average measurement and the ground truth, where the difference is normalized by the standard error across measurements. Variance is quantified as the median standard deviation across measurements. Error is quantified as the correlation between each measurement and the ground truth, averaged across measurements. The gray vertical line indicates the bias value associated with the case of unbiased measurement (assuming Gaussian noise).

In the second simulation, we use as ground truth a synthetic hemodynamic response function (**Figure 3A, top**) and simulate noisy measurements of this ground truth by adding temporally correlated Gaussian noise. As expected, the raw data ('No denoising') follow the ground truth, in the sense of lacking bias, but suffer from high variance (**Figure 3B, first column**). The method of reconstructing the measurements using a small set of basis functions ('Basis restriction') greatly reduces variance but incurs major deviations from the ground truth (**Figure 3B, second column**). The discrepancy can be traced to the fact that the basis functions do not have much dynamics around the time of the undershoot (see blue arrow).

The method of fitting a parametric function to the data ('Parametric fit') provides variance reduction and, appealingly, introduces very little bias, if any (**Figure 3C, third column**). This makes sense, since the parametric function used to fit the data is the same function that was used to generate the ground truth. If a different parametric function were used, these results of course may no longer hold.

The quantitative summary plots (**Figure 3C**) bear out the above observations. Basis restriction is very effective at reducing variance but is highly biased (location 1). Nonetheless, on balance, the bias-variance tradeoff is such that error is reduced compared to no denoising (location 2). However, there is even a better method: parametric fitting is essentially unbiased (location 3) and performs the best at achieving results that are similar to the ground truth (location 4). Interestingly, even though parametric fitting has *more* variance across analysis results than basis restriction, parametric fitting yields results that better match ground truth (arrow 5). This can be understood as the consequence of the undesirable bias that is induced by basis restriction.

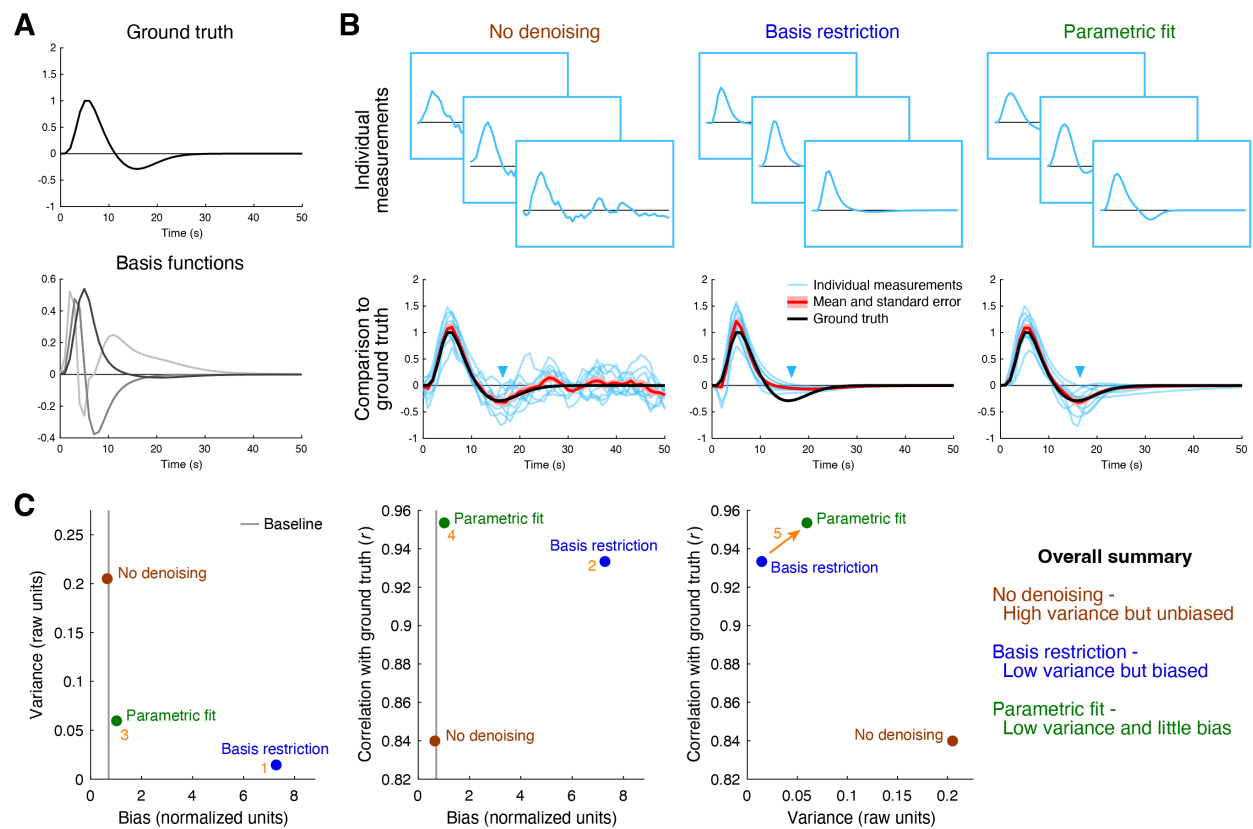

**Figure 3. Denoising response timecourses.** In this simulation (code available at https://osf.io/6jhmr/), we generate noisy measurements by starting with a ground-truth hemodynamic response function (HRF) and adding temporally correlated Gaussian noise. We then attempt to denoise the data using different denoising methods: no denoising, reconstruction using a restricted set of basis functions, fitting using a parametric model. *A*, Reference timecourses. We illustrate the ground-truth HRF (top) and the temporal basis functions used in one of the denoising methods (bottom). *B*, Denoising results. Each column shows results for a different denoising method. We show three example measurements (top row) and comparison to the ground truth (bottom row). *C*, Quantitative assessment of bias, variance, and error. Same format as **Figure 2C**.

In the third simulation, we use as ground truth a synthetic set of tuning curves (10 units, 50 conditions) whose dimensionality is fixed to 4 (**Figure 4A**) and simulate noisy measurements of this ground truth by adding Gaussian noise. As expected, the raw data ('No denoising') follow the ground truth, in the sense of lacking bias, but suffer from high variance (**Figure 4B, first column**). The method of boxcar smoothing substantially reduces variance and, appealingly, does not incur any appreciable bias (**Figure 4B, second column**). This makes sense given that the width of the boxcar used is 3, which is relatively small compared to the intrinsic smoothness of the ground-truth tuning curves. The method of dimensionality reduction using principal components analysis (PCA) yields variance reduction at the expense of bias, with the specific bias-variance tradeoff controlled by the number of dimensions. Specifically, if dimensionality is aggressively reduced, more variance reduction is achieved but more bias is introduced (e.g., **Figure 4B, sixth column**). If dimensionality is reduced less aggressively, less variance reduction is achieved but less bias is introduced (e.g., **Figure 4B, third column**).

The quantitative summary plots (**Figure 4C**) provide additional insight. The bias-variance tradeoff in PCA is clearly visible: there is a continuous progression from PCA6 to PCA2 in terms of increasing amounts of bias and decreasing amounts of variance (arrow 1). Compared to PCA8, PCA6 does not incur appreciable bias; this suggests that preserving six dimensions is sufficient to retain all (or nearly all) of the underlying signal in the noisy measurements. The method of boxcar smoothing clearly outperforms PCA, as it greatly reduces variance and does not induce bias (location 2), and, moreover, achieves the best match to ground truth (location 3). Interestingly, the number of dimensions in PCA that maximizes similarity to ground truth is 3 (location 4), which is not the same as the true dimensionality of the underlying representation. This may seem counterintuitive at first, but can be understood as the simple consequence of the mixing of bias and variance when quantifying similarity to ground truth. That is, even though retaining only 3 dimensions is guaranteed to discard some of the true signal and incur bias (since the ground-truth dimensionality is 4), the reduction of variance afforded by retaining only 3 dimensions apparently improves the overall similarity to ground truth. Perhaps the most important insight is that reducing dimensionality to 4 already starts to introduce noticeable levels of bias (location 5). This is due to the fact that in the presence of measurement noise, the dimensions identified by PCA will start to deviate from the true dimensions that underlie the ground-truth representation. In other words, noise inevitably corrupts all of the PCA-identified dimensions, not just the ones that are discarded (Veraart et al., 2016). Hence, there is no guarantee that using 4 dimensions will retain all of the relevant signal contained in a given measurement.

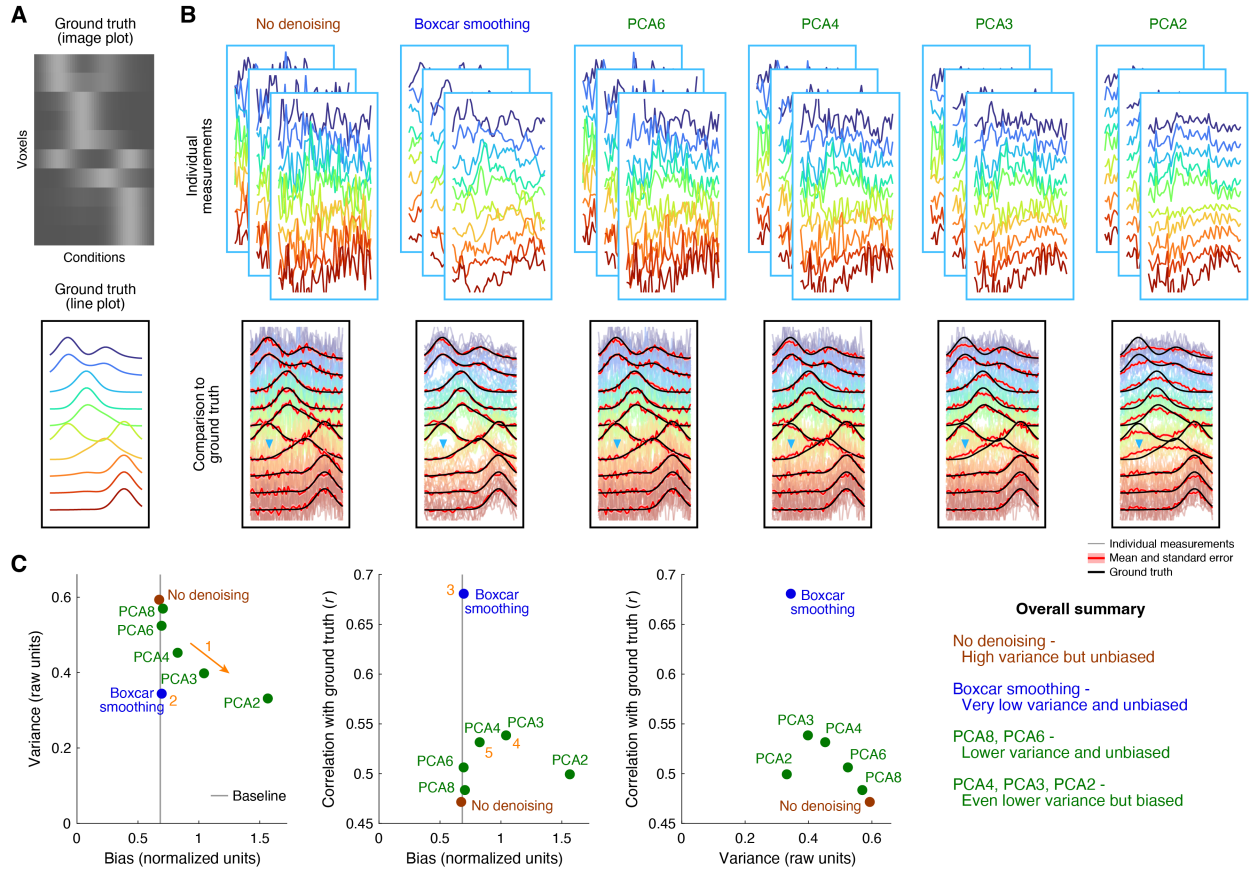

**Figure 4. Denoising tuning curves.** In this simulation (code available at https://osf.io/a6k9m/), we generate noisy measurements by starting with a ground-truth collection of tuning curves whose underlying dimensionality is fixed at 4 and adding Gaussian noise independently to each data point. We then attempt to denoise the data using different denoising methods: no denoising, simple boxcar smoothing of responses to nearby conditions, and dimensionality reduction using principal component analysis (PCA). *A*, Reference data. We illustrate the ground-truth tuning curves as an image (top) and as line plots (bottom). Color is used to distinguish different units. *B*, Denoising results. Each column shows results for a different denoising method. We show three example measurements (top row) and comparison to the ground truth (bottom row). *C*, Quantitative assessment of bias, variance, and error. Same format as **Figure 2C**.

# Discussion

In this paper, we have described a simple framework for evaluating denoising methods, and we have provided examples that highlight important (and possibly surprising) observations about denoising. These examples were not intended to benchmark the performance of state-of-the-art methods, but rather to demonstrate insights into the nature of denoising. The main issues that we focus on—bias and variance—are well-understood in statistics. We believe these issues need increased attention in experimental fields, especially in light of the increasing complexity of datasets and analysis pipelines. While developing denoising techniques to improve data quality is a worthwhile endeavor, we should approach such techniques with caution and strive to avoid introducing systematic bias to our measurements (see also the perspective by Hoffman et al., 2021). To summarize our viewpoint, we propose the following three action items.

## We should acknowledge bias

As a first action item, we should acknowledge bias as a major potential concern when applying denoising methods. When making measurements, a presumption is that repeated measurements will help the researcher narrow the range of plausible values for the parameter of interest. In this context, systematic bias should be alarming. Some denoising methods might not introduce bias, and it might be possible to see that this is the case from a theoretical perspective. However, in general, denoising methods are likely bound to the bias-variance tradeoff: there is likely going to be a tradeoff between reduction of variance and introduction of bias. Even if one does not yet know exactly what the bias is for a given method, it is worthwhile to acknowledge and discuss what this potential bias might be.

An intuitive way to think about bias is through the concept of a prior. Denoising methods can be viewed as bringing priors to a set of data (Knoll et al., 2020). On the one hand, if we do not incorporate any priors, the data in their raw form are noisy but safe: they can be expected to provide the right answer on average (assuming that the noise is zero-mean). On the other hand, if we apply a denoising method, we are bringing in priors, or implicit assumptions, regarding the nature of the underlying system. The key question is whether the priors embodied by the denoising method are a good match to the system. If the priors are very well matched (e.g. **Figure 2B, fourth column**), little or no bias is introduced, and we can enjoy the reduced variance. If the priors are not well matched (e.g. **Figure 3B, second column**), bias is introduced, and the reduced variance may not be worth it. Whether a given denoising method is well matched to a system may vary across situations. For example, anisotropic smoothing (e.g. **Figure 2B, fourth column**) is likely inappropriate for structures consisting of point-like features; Gaussian smoothing (e.g. **Figure 2B, second column**) is actually a good approach in situations where the measurement resolution is high compared to the scale of the underlying signal of interest. Analogous to the "No Free Lunch Theorem" in optimization (Wolpert and Macready, 1997), we should recognize that any given denoising method is not guaranteed to perform well in all situations. Accordingly, our goal should not only be to demonstrate that a given denoising method performs well in certain situations, but should also be to identify the range of situations within which the method performs well and the range within which the method fails.

Illustrative examples of priors come in cases where there are literally no data. These cases conveniently expose the full nature of the prior embodied by a technique. For instance, suppose we delete a small region of a photograph and use an image inpainting technique to fill in the region. While we are likely to obtain a reasonable-looking image that generally conforms to natural image statistics, it is obvious that this is no substitute for actual measurement. Had there been a specific object in the deleted region, it is likely that the inpainting technique would miss this completely and instead fill the region with general texture priors (Li, 2020). In other words, the technique would likely incur massive bias. Or, as a different example, suppose we train a model to predict high-resolution details that typically accompany low-resolution measurements. This model might be quite effective within a certain data regime at predicting high-resolution details when only low-resolution data are available, but might make non-sensical predictions when exposed to novel data regimes that differ substantially from the training dataset (Hoffman et al., 2021).

## We should study and quantify bias

As a second action item, we should study the bias that may be present in a denoising technique, and quantify its magnitude in real-world situations. Carefully characterizing the bias of a method is useful for providing full transparency and enabling accurate risk assessment. Bias can be studied using different approaches. It might be possible to make a theoretical assessment as to whether a denoising technique is likely to incur bias and what this bias might be like. This is more feasible for denoising techniques that are based on simple, clear principles. For example, although simple smoothing is a naive approach, one appealing feature is that we fully understand the risk of bias that it entails. In contrast, denoising techniques that derive their power from large amounts of training data (e.g. deep neural networks) or techniques that derive noise estimates from the data themselves are more difficult to assess from an *a priori* perspective. Alternatively, the bias in denoising techniques can be evaluated empirically (like the examples shown in **Figures 2–4**).

One of the take-home points of this paper is that not all denoising performance metrics are equally useful or informative:

- *Good* is quantifying variance. Examining variability of results across repeated measurements or independent splits of a dataset certainly provides useful information. However, as noted earlier, reliability does not provide information about potential bias.
- *Better* is quantifying error. This is a widely used approach in image processing (Chatterjee and Milanfar, 2009; Zbontar et al., 2018) in which one seeks to minimize the error between a denoised output and a reference ground-truth image. Error can be quantified in various ways, such as mean squared error, peak signal-to-noise ratio, or structural similarity index (Fan et al., 2019; Wang et al., 2004). Alternatively, error can be assessed through the use of cross-validation to assess generalization to unseen samples, which serve as an implicit ground truth. Although the metric of error is sensitive to bias, it is also affected by variance and therefore provides ambiguous information. In other words, error is insufficient to understand whether and to what extent a method suffers from bias.
- *Best* is quantifying bias separately from variance and error. This can be done by applying a denoising method to multiple independent measurements and carefully comparing the mean of the results to a ground-truth measure (as shown in **Figures 2–4**).

Overall, we stress the importance of assessing both bias and variance. One approach to adjudicating amongst different denoising methods is to establish a data regime for which we would like a method to perform well, determine which methods are unbiased (or nearly unbiased) in this regime, and then select from these methods the one that has the least variance. We make these suggestions under full acknowledgement that we ourselves have not fully implemented these ideas in the past. For example, we used cross-validation to evaluate denoising performance in this study (Kay et al., 2013), but it would have been even more informative had we specifically assessed bias.

Although we demonstrate ground-truth simulations in this paper, studying bias is not limited to such situations. On the one hand, ground-truth simulations can deliver many valuable insights (Huang et al., 2021; Lerma-Usabiaga et al., 2020). However, ground-truth simulations are susceptible to the criticism that they may not capture the full complexity of real empirical data. Fortunately, it is possible to study bias in real data if one has access to a dataset in which many repeated measurements are available. One approach is to average across these measurements, treat the result as ground truth, and evaluate how well denoising methods can use single (or a few) measurements to recover the ground truth. Note that perfect recovery is not necessarily desired in this scenario since the ground-truth measure is still subject to some amount of measurement error.

Denoising efforts, especially in the field of machine learning, often place great emphasis on improving predictive performance, in the sense of generating results that better approximate a target ground-truth measure. While this engineering mindset has obvious practical and commercial value (Efron and Hastie, 2021) and can be quite effective in driving competition and therefore progress (Donoho, 2017), it falls short as a means for assessing measurement accuracy. Specifically, *predictive performance reflects a combination of bias and variance and therefore is insufficient in and of itself for studying bias*. Unless predictive performance is perfect, there is a potential that bias exists for a given denoising technique. Emphasis on prediction can be viewed in terms of the divide between what has been termed 'predictive modeling' and 'explanatory modeling': "predictive modeling seeks to minimize the combination of bias and estimation variance, occasionally sacrificing theoretical accuracy [i.e. correct identification of properties of the underlying system] for improved empirical precision" (Shmueli, 2010).

## We should consider the risk of bias to one's goals

As a third action item, we should consider the risk of bias in the context of the broader goals of a given endeavor. All else being equal, we would argue that for everyday scientific measurements, we cannot risk using denoising methods that introduce bias, as this may lead to incorrect inferences from the data. However, adopting a more realistic perspective, we recognize that the bias that might be present in a given denoised dataset could be a relatively minor aspect of the data. Even if we know with certainty that a denoising method introduces bias, we might reasonably ask: how strongly does the bias affect the main issues at stake?

A reasonable strategy is to consider each situation on a case-by-case basis and make a deliberate decision regarding the risk of bias. In some situations, bias might be acceptable. For example, if the goal is to clean an audio or video signal for aesthetic purposes or for basic perceptual interpretation (Fan et al., 2019), then bias would seem to cause little harm and a reasonable stance is to simply resign and accept bias (as in Chatterjee and Milanfar, 2009). Or, if a denoising method affects an aspect of a given dataset that does not have substantial impact on the main findings from the data (e.g., small changes in region identification might be unlikely to change the overall measured activity from a brain region), then bias would not seem to be a problem. In other situations, bias might be unappealing but must be accepted out of necessity. For example, if a dataset is too noisy to make inferences and additional measurements are not possible (e.g., due to the rarity of the data), it may be necessary to apply a denoising method in order to salvage the data and make some inferences, even if imperfect. However, there are certainly situations where bias may be unacceptable. For example, if a set of noisy data is being used to make a clinical diagnosis, it might be better to leave the data untouched and acknowledge that the data are inconclusive than to risk introducing an artifact or removing a true signal. Or, as another example, if a set of data is intended to critically test hypotheses about temporal characteristics of a system, one might avoid applying a denoising method that has access to multiple temporal measurements, as the method might potentially introduce biases in the temporal domain, and instead restrict the method to single measurements at a time. In general, domain expertise and deep understanding of a denoising method are required in order to make an informed decision regarding the acceptability of bias.

## Author Contributions

K.K. implemented the simulations and wrote the paper.

## Acknowledgements

We thank M. Akcakaya, O. Gulban, E. Merriam, A. Rokem, and J. Winawer for helpful discussions and comments on the manuscript.

## Competing Interests

The author confirms that there are no competing interests.

## Data and code availability statement

The data and code used to carry out the simulations in this paper are available at https://osf.io/weg87/. These materials can be easily adapted to explore different simulation scenarios (different noise levels, different ground truths, etc.).

# References

Abbott, A., 2021. How the world's biggest brain maps could transform neuroscience. Nature 598, 22–25.
Allen, E.J., St-Yves, G., Wu, Y., Breedlove, J.L., Prince, J.S., Dowdle, L.T., Nau, M., Caron, B., Pestilli, F., Charest, I., Hutchinson, J.B., Naselaris, T., Kay, K., 2022. A massive 7T fMRI dataset to bridge cognitive neuroscience and artificial intelligence. Nat. Neurosci. 25, 116–126.
Behzadi, Y., Restom, K., Liau, J., Liu, T.T., 2007. A component based noise correction method (CompCor) for BOLD and perfusion based fMRI. Neuroimage 37, 90–101.
Blei, D.M., Smyth, P., 2017. Science and data science. Proc. Natl. Acad. Sci. U. S. A. 114, 8689–8692.
Chatterjee, P., Milanfar, P., 2009. Is denoising dead? IEEE Trans. Image Process.
Deng, J., Dong, W., Socher, R., Li, L., Kai Li, Li Fei-Fei, 2009. ImageNet: A large-scale hierarchical image database. Presented at the 2009 IEEE Conference on Computer Vision and Pattern Recognition.
Donoho, D., 2017. 50 Years of Data Science. J. Comput. Graph. Stat. 26, 745–766.
Efron, B., Hastie, T., 2021. Computer age statistical inference, student edition, Institute of Mathematical Statistics Monographs. Cambridge University Press, Cambridge, England.
Fadnavis, S., Batson, J., Garyfallidis, E., 2020. Patch2Self: Denoising Diffusion MRI with Self-Supervised Learning. arXiv [cs.LG].
Fan, L., Zhang, F., Fan, H., Zhang, C., 2019. Brief review of image denoising techniques. Vis Comput Ind Biomed Art 2, 7.
Gavish, M., Donoho, D.L., 2017. Optimal Shrinkage of Singular Values. IEEE Trans. Inf. Theory 63, 2137–2152.
Gulban, O.F., Schneider, M., Marquardt, I., Haast, R.A.M., Martino, F.D., 2018. A scalable method to improve gray matter segmentation at ultra high field MRI. PLoS One 13, e0198335.
Hastie, T., Tibshirani, R., Friedman, J.H., 2001. The elements of statistical learning: data mining, inference, and prediction, Springer series in statistics. Springer, New York.
Hoffman, D.P., Slavitt, I., Fitzpatrick, C.A., 2021. The promise and peril of deep learning in microscopy. Nat. Methods.
Huang, P., Correia, M.M., Rua, C., Rodgers, C.T., Henson, R.N., Carlin, J.D., 2021. Correcting for Superficial Bias in 7T Gradient Echo fMRI. Front. Neurosci. 15, 715549.
Kay, K.N., Rokem, A., Winawer, J., Dougherty, R.F., Wandell, B., 2013. GLMdenoise: a fast, automated technique for denoising task-based fMRI data. Front. Neurosci. 7, 247.
Knoll, F., Hammernik, K., Zhang, C., Moeller, S., Pock, T., Sodickson, D.K., Akçakaya, M., 2020. Deep-Learning Methods for Parallel Magnetic Resonance Imaging Reconstruction: A Survey of the Current Approaches, Trends, and Issues. IEEE Signal Process. Mag. 37, 128–140.
Lecoq, J., Oliver, M., Siegle, J.H., Orlova, N., Ledochowitsch, P., Koch, C., 2021. Removing independent noise in systems neuroscience data using DeepInterpolation. Nat. Methods 18, 1401–1408.
LeCun, Y., Bengio, Y., Hinton, G., 2015. Deep learning. Nature 521, 436–444.
Lerma-Usabiaga, G., Benson, N., Winawer, J., Wandell, B.A., 2020. A validation framework for neuroimaging software: The case of population receptive fields. PLoS Comput. Biol. 16, e1007924.
Li, C.-T., 2020. A Practical Generative Deep Image Inpainting Approach. Towards Data Science. URL https://towardsdatascience.com/a-practical-generative-deep-image-inpainting-approach-1c99fef68bd7
Marx, V., 2013. Biology: The big challenges of big data. Nature 498, 255–260.
Mason, H.T., Graedel, N.N., Miller, K.L., Chiew, M., 2021. Subspace-constrained approaches to low-rank fMRI acceleration. Neuroimage 238, 118235.
Pruim, R.H.R., Mennes, M., van Rooij, D., Llera, A., Buitelaar, J.K., Beckmann, C.F., 2015. ICA-AROMA: A robust ICA-based strategy for removing motion artifacts from fMRI data. Neuroimage 112, 267–277.
Qiao, C., Li, Di, Guo, Y., Liu, C., Jiang, T., Dai, Q., Li, Dong, 2021. Evaluation and development of deep neural networks for image super-resolution in optical microscopy. Nat. Methods 18, 194–202.
Raunig, D.L., McShane, L.M., Pennello, G., Gatsonis, C., Carson, P.L., Voyvodic, J.T., Wahl, R.L., Kurland, B.F., Schwarz, A.J., Gönen, M., Zahlmann, G., Kondratovich, M.V., O'Donnell, K., Petrick, N., Cole, P.E., Garra, B., Sullivan, D.C., QIBA Technical Performance Working Group,

  2015. Quantitative imaging biomarkers: a review of statistical methods for technical performance assessment. Stat. Methods Med. Res. 24, 27–67.

Salimi-Khorshidi, G., Douaud, G., Beckmann, C.F., Glasser, M.F., Griffanti, L., Smith, S.M., 2014. Automatic denoising of functional MRI data: combining independent component analysis and hierarchical fusion of classifiers. Neuroimage 90, 449–468.

Shmueli, G., 2010. To explain or to predict? Stat. Sci. 25, 289–310.

Veraart, J., Novikov, D.S., Christiaens, D., Ades-Aron, B., Sijbers, J., Fieremans, E., 2016. Denoising of diffusion MRI using random matrix theory. Neuroimage 142, 394–406.

Vizioli, L., Moeller, S., Dowdle, L., Akçakaya, M., De Martino, F., Yacoub, E., Uğurbil, K., 2021. Lowering the thermal noise barrier in functional brain mapping with magnetic resonance imaging. Nat. Commun. 12, 5181.

Wang, Z., Bovik, A.C., Sheikh, H.R., Simoncelli, E.P., 2004. Image quality assessment: from error visibility to structural similarity. IEEE Trans. Image Process. 13, 600–612.

Weickert, J., 1998. Anisotropic diffusion in image processing. Teubner Stuttgart.

Wolpert, D.H., Macready, W.G., 1997. No free lunch theorems for optimization. IEEE Trans. Evol. Comput. 1, 67–82.

Yang, Z., Zhuang, X., Sreenivasan, K., Mishra, V., Cordes, D., Alzheimer's Disease Neuroimaging Initiative, 2020. Disentangling time series between brain tissues improves fMRI data quality using a time-dependent deep neural network. Neuroimage 223, 117340.

Zbontar, J., Knoll, F., Sriram, A., Murrell, T., Huang, Z., Muckley, M.J., Defazio, A., Stern, R., Johnson, P., Bruno, M., Parente, M., Geras, K.J., Katsnelson, J., Chandarana, H., Zhang, Z., Drozdzal, M., Romero, A., Rabbat, M., Vincent, P., Yakubova, N., Pinkerton, J., Wang, D., Owens, E., Lawrence Zitnick, C., Recht, M.P., Sodickson, D.K., Lui, Y.W., 2018. fastMRI: An Open Dataset and Benchmarks for Accelerated MRI. arXiv [cs.CV].